\documentclass[prd,aps,twocolumn,leterpaper,preprintnumbers]{revtex4}
\usepackage{amsmath,amssymb}
\usepackage{eucal}
\usepackage[dviwindo]{graphicx}

\newcommand{\bs}{\begin{subequations}}
\newcommand{\es}{\end{subequations}}
\numberwithin{equation}{section}
\newcommand{\ben}{\begin{eqnarray}}
\newcommand{\een}{\end{eqnarray}}
\newcommand{\la}{\label}
\begin{document}

\title{Teukolsky-Starobinsky Identities -- a Novel Derivation and Generalizations}

\author{Plamen~P.~Fiziev
\footnote{E-mail:\,\,\,fiziev@phys.uni-sofia.bg}}

\affiliation{Department of Theoretical Physics, University of Sofia, Boulevard
5 James Bourchier, Sofia 1164, Bulgaria
and BLTF, JINR, Dubna, 141980 Moscow Region, Rusia}

\begin{abstract}
We present a novel derivation of the Teukolsky-Starobinsky identities,
based on properties of the confluent Heun functions. These functions define analytically
all exact solutions to the Teukolsky master equation, as well as to the Regge-Wheeler and Zerilli ones.
The class of solutions, subject to Teukolsky-Starobinsky type of identities is studied.
Our generalization of the Teukolsky-Starobinsky identities is valid for the already studied
linear perturbations to the Kerr and Schwarzschild metrics, as well as for
large new classes of of such perturbations which are explicitly described in the present article.
Symmetry of parameters of confluent Heun's functions is shown to stay behind the
behavior of the known solutions under the change of the sign of their spin weights.
A new efficient recurrent method for calculation of  Starobinsky's constant is
described.

\end{abstract}
\sloppy
\maketitle
\section{Introduction}

Today the Teukolsky-Starobinsky identities (TSI)
are essential ingredient of the theory of
perturbations to a gravitational field of Kerr black holes.
For spin one and two the TSI were discovered together with their first applications
in the study of perturbations of rotating relativistic objects
pioneered by Teukolsky \cite{Teukolsky} and Starobinsky \cite{Starobinsky}.
An independent derivation and detailed study of TSI was given by Chandrasekhar in
\cite{Chandra,ChandraMT} and by other authors in \cite{other}.
Later on a proper generalization of the TSI for all physically interesting spins,
as well as the TSI in the presence of a nonzero cosmological constant
was found and justified in \cite{TSI}.
Some more recent applications can be found in \cite{TSIapplications}.

All known derivations of TSI are based on a direct use of the famous
Teukolsky master equation (TME), or its generalizations,
and on some very special properties of the solutions in use. For example,
one utilizes the regularity and integrability of the used angular solutions,
parameters with real values, etc.

The TME describes the perturbations
$\Psi(t,r,\theta,\varphi)$ of fields of all interesting spin weights $s=0,
\pm 1/2, \pm 1, \pm 3/2, \pm 2$ in Kerr background in terms
of Newman-Penrose scalars. Various significant results and
references may be found in
\cite{Teukolsky,Starobinsky,Chandra,ChandraMT,other,TSI,TSIapplications,QNM}.

The key feature is that in Boyer-Lindquist coordinates
one can separate the variables using the ansatz
$\Psi(t,r,\theta,\varphi)=\exp(-i\omega t + im
\varphi)R(r)S(\theta)$.
Thus, a pair of two connected differential equations arises -- the Teukolsky
angular equation (TAE):
\ben {\frac {1}{\sin \theta}} {\frac {d}{d\theta}} \left( \sin
\theta {\frac {d} {d\theta}}S( \theta ) \right)
+\Big(W(\theta)+E\Big) S(\theta)\!=0, \hskip 0truecm \la{angE} \een
$W(\theta)\!=\!a^2\omega^2\!\cos^{2}\! \theta\!-\!2s a\omega \cos
\theta\! -\!(m^2\!+\!s^2\!+\!2ms\cos\theta)/\sin^2 \theta $, and the Teukolsky
radial equation (TRE):
\ben {\Delta}^{-s}{\frac {d }{dr}} \left( {\Delta}^{s+1}{ \frac
{d}{dr}}R(r)  \right) + \Big( V(r)-E \Big) R(r) =0, \la{radE} \een
$V(r)\!=\!4is\omega r+2 m a\omega-a^2\omega^2+s(s\!+\!1)\!+\!{\frac{K^2-2is\left(r-M\right)
K}\Delta}$.  The real parameter $a=J/M\geq 0$
is related with angular momentum $J$ of the Kerr metric
($a<M$ for black holes, or $a>M$ -- for naked singularities),
$M$ is the Keplerian mass of the Kerr solution, $K\!=\!\omega(r^2\!+\!a^2)\!-\!ma$,
$\Delta\!=\!(r\!-\!r_{+})(r\!-\!r_{-})\!=\!r^2\!-\!2Mr\!+\!a^2$.
The two {\em complex} separation constants $\omega$ and $E$ are
to be determined using the boundary conditions of the problem.

At present the usage of the Kerr metric for the description of astrophysical black holes is widely
accepted. The astrophysical application of the naked singularities seems to be more problematic
because of their instability \cite{VCMC}.
However, such instability may be useful for construction of models
of astrophysical explosions like gamma ray bursts \cite{STARA}.
Our method for the derivation of the generalized TSI is applicable to both cases,
despite of the fact that some quantities (like $r_\pm$)
are complex for the naked singularities.

The TSI for angular Teukolsky's function  ${}_{s}S(\theta)$
of arbitrary spin weight $s$ can be written in the form
\ben
\la{TSI_A}
{}_{1-s}\mathcal{L}\,\,\,{}_{2-s}\mathcal{L}\,\,\dots{}_{s-1}\mathcal{L}\,\,\,{}_{s}\mathcal{L}\,\big({}_{\!+s} S(\theta)\big)&=& {}_{s}\mathfrak{C}\,\big({}_{\!-s}S(\theta)\big),\hskip 1truecm\\
{}_{1-s}\mathcal{L}^{\!\dag}\,{}_{2-s}\mathcal{L}^{\!\dag}\,\dots{}_{s-1}\mathcal{L}^{\!\dag}\,{}_{s}\mathcal{L}^{\!\dag}
\,\big({}_{\!-s}S(\theta)\big)&=&{}_{s}\mathfrak{C}\,\big({}_{\!+s}S(\theta)\big);
\nonumber
\een
where
\ben
{}_{n}\mathcal{L}=\partial_\theta+\left({m\over{\sin\theta}}-a\omega\sin\theta\right)+n\cot\theta,
\\
{}_{n}\mathcal{L}^{\!\dag}=\partial_\theta-\left({m\over{\sin\theta}}-a\omega\sin\theta\right)+n\cot\theta.
\nonumber
\een

Using the operators
\ben
\la{TSI_RD}
{}_n\mathcal{D}=\partial_r+i\,{\frac K \Delta}+{\frac n \Delta}\partial_r\Delta,\\
{}_n\mathcal{D}^{\dag}=\partial_r-i\,{\frac K \Delta}+{\frac n \Delta}\partial_r\Delta,\nonumber
\een
one obtains the TSI for the radial Teukolsky function  ${}_{s}R(r)$ of arbitrary spin weight $s$
in the form \footnote{Note that we are using Teukolsky's convention for the signs
of the frequency $\omega$ and $K$. It is opposite to Chandrasekhar's one.
Our conventions for notation ${}_n\mathcal{D}$ and ${}_n\mathcal{D}^{\dag}$,
as well as for the phase of Starobinsky's constant ${}_{s}\mathfrak{D}$
are also opposite with respect to the standard conventions.
This way we obtain the form \eqref{TSI_R} of the radial TSI
which is in accord to the form \eqref{TSI_A} of the angular TSI
with respect to the places of the signs '$\pm$' of the spin weights $\pm s$
in the corresponding Teukolsky's functions.}:
\ben
\la{TSI_R}
\Delta^s\,\left({}_0\mathcal{D}\right)^{2s}\big(\Delta^s\,{}_{+s}R(r)\big)&=&
{}_{s}\mathfrak{D}\,\big({}_{-s}R(r)\big),\\
\Delta^s\,\left({}_0\mathcal{D}^\dag\right)^{2s}\big({}_{-s}R(r)\big)&=&
{}_{s}\mathfrak{D}^*\!\big(\Delta^s\,{}_{+s}R(r)\big).\nonumber
\een

In Eqs. \eqref{TSI_A} and \eqref{TSI_R} ${}_{s}\mathfrak{C}$ and ${}_{s}\mathfrak{D}$
are the corresponding Starobinsky constants \footnote{In \eqref{TSI_R} the star ${}^*$ denotes a complex conjugation.}.
The standard calculation of these constants for arbitrary $s$
is quite complicated \cite{Chandra,ChandraMT}.

In the widespread derivation of TSI, proposed for the first time in \cite{Chandra},
one supposes to work with {\em real} frequencies $\omega$.
This is explicitly stressed in \cite{Chandra}, as well as in some of the articles \cite{other,TSI}.
As a result, the operators $\mathcal{D}$ and $\mathcal{D}^\dag$ can be considered conjugated;
this property is systematically used for derivation of the TSI \cite{Chandra,ChandraMT,other,TSI}.
However, the  well-known boundary problems,
related to black holes and naked singularities, yield complex frequencies $\omega$
with certainly nonzero imaginary part $\Im(\omega)\neq 0$
(See for example the Refs. \cite{ChandraMT,QNM} and the recent articles \cite{PFDS},
where a new boundary problem was studied and some very preliminary
results of our attempts to apply singular solutions of TAE to the description
of collimated relativistic jets are reported.)
The physical reason is clear:
The systems under consideration are open physical systems
and their energy is not conserved. Hence, for them $\Im(\omega)\neq 0$.
Our derivation of TSI does not imply complex conjugation of the corresponding quantities.
It is valid for complex frequencies $\omega$,
as well as for complex values of all other parameters.
Despite the fact that our derivation is quite different in form,
it is closely related to the original one, outlined in \cite{Teukolsky,Starobinsky}.
This original method is also valid for complex frequencies.

It has been well known for a long time \cite{TME_Heun} that for $a\neq M$ \footnote{Here
we do not consider the extremal Kerr black holes with $a=M$,
since for them the corresponding solutions are not described by the confluent
Heun functions. Instead, one must use
bi-confluent Heun's functions with quite different properties
\cite{Heun}. Therefore this case needs a separate treatment.
For application of
the bi-confluent Heun functions for description of perturbations of Kasner spacetimes
see  R. Pons and G. Marcilhacy, Class. Quantum Grav. {\bf 4}, 171-179 (1987).}
the TAE (\ref{angE}) and TRE (\ref{radE}) can be reduced to the confluent Heun ordinary
differential equation \cite{Heun}.
Recently in \cite{Fiziev1} all classes of the exact solutions to
TAE (\ref{angE}) and  TRE (\ref{radE}) were described using
the confluent Heun function $\text{HeunC}(\alpha,\beta,\gamma,\delta,\eta,z)$
-- a unique particular solution of the confluent Heun equation,
written here in the simplest uniform shape:
\ben
H''\!+\!\left(\alpha\!+\!{\frac{\beta\!+\!1}{z}}\!+\!{\frac{\gamma\!+\!1}{z\!-\!1}}\right)H'\!+\!
\left({\frac\mu z}\!+\!{\frac\nu {z\!-\!1}} \right)H\!=\!0.\la{HeunCDE}
\een
The function $\text{HeunC}(\alpha,\beta,\gamma,\delta,\eta,z)$
is defined as a solution of Eq. \eqref{HeunCDE} which is regular in vicinity of
the singular point $z\!=\!0$ and
subject to the normalization condition $\text{HeunC}(\alpha,\beta,\gamma,\delta,\eta,0)=1$ \cite{Heun}.
The parameters $\alpha, \beta, \gamma, \delta, \eta $, introduced in
\cite{Heun} and used in the widespread computer package \textsc{MAPLE},
are related with $\mu$ and $\nu$, according to the equations
$\mu\!=\!{\frac{1} 2}(\alpha-\beta -\gamma+\alpha\beta-\beta\gamma)-\eta$ and
$\nu\!=\!{\frac{1} 2}(\alpha+\beta +\gamma+\alpha\gamma+\beta\gamma)+\delta + \eta$.

The other particular solutions to Eq. (\ref{HeunCDE})
are {\em not} termed "confluent Heun's functions",
according to the accepted modern terminology \cite{Heun}.
The reason is that in the general case other solutions can be represented in a
nontrivial way in terms of solutions $\text{HeunC}(\alpha,\beta,\gamma,\delta,\eta,z)$.
An example if the application of this specific property of the Eq. \eqref{HeunCDE}
is the basic formula \eqref{X} in the next section.
Therefore from a computational point of view it is sufficient to study
the Taylor series  of this standard local solution
and its analytical continuation in the complex plane $\mathbb{C}_z$.
Thus, the instrumental use of the confluent Heun function
$\text{HeunC}(\alpha,\beta,\gamma,\delta,\eta,z)$
is much more advantageous than the simple fact, recognized already in \cite{TME_Heun},
that the TRE and TAE can be reduced to the Eq. \eqref{HeunCDE}.

Essential novel properties of the confluent Heun functions
$\text{HeunC}(\alpha,\beta,\gamma,\delta,\eta,z)$ were
studied for the first time in the recent article \cite{Fiziev2}.
These properties are prerequisites for our derivation of the TSI in the next section.
The subclass of $\delta_N$-confluent-Heun's-functions
\ben
\la{HeunCN}
\text{HeunC}_N(\alpha,\beta,\gamma,\eta,z)\!=\!
\text{HeunC}\big(\alpha,\beta,\gamma,\delta_{N},\eta,z\big),\hskip .3truecm
\een
was introduced assuming the fulfilment of the "$\delta_N$-condition,"
\ben
\delta_{N}\!=\!-\alpha\left(N\!+\!1\!+\!(\beta\!+\!\gamma)/2\right),
\la{deltaN}
\een
for some fixed nonnegative integer $N=0, 1, 2,\dots$

For functions (\ref{HeunCN}) and their associate functions
\ben
\la{AHeunCN}
\text{HeunC}^\maltese_N(\alpha,\beta,\gamma,\eta,z)=\hskip 3.9truecm \\
=\text{HeunC}\big(\alpha_{N}^{{}_\maltese},\beta_{N}^{{}_\maltese},\gamma_{N}^{{}_\maltese},
\delta_{N}^{{}_\maltese},\eta_{N}^{{}_\maltese},z\big),
\hskip .2truecm \nonumber
\een
\ben
\la{AHeunCNparameters}
\alpha_{N}^{{}_\maltese}\!=\!\alpha,\,\beta_{N}^{{}_\maltese}\!=
\!\beta\!+\!N\!+\!1,\,\gamma_{N}^{{}_\maltese}\!=\!\gamma\!+\!N\!+\!1,
\hskip 1.2truecm \\
{\frac{\delta_{N}^{{}_\maltese}}{\alpha_{N}^{{}_\maltese}}}\!=\!{\frac{\delta}{\alpha}}\!+\!N\!+\!1,\,
\eta_{N}^{{}_\maltese}\!=\!\eta\!+\!{\frac {(N\!+\!1)(\!N\!+\!1\!-\!\varkappa)} 2},
\hskip .4truecm \nonumber
\een
$\varkappa\!=\!\alpha\!-\!\beta\!-\!\gamma$; the following simple basic relation is available:
\ben
\la{HHn_maltese}
{\frac{d^{N\!+1} }{dz^{N\!+1}}}\,\text{HeunC}_N(\alpha,\beta,\gamma,\eta,z)=\hskip 1.7truecm\\
=\mathfrak{P}_N\, \text{HeunC}^\maltese_N(\alpha,\beta,\gamma,\eta,z).
\hskip .4truecm\nonumber
\een
$\mathfrak{P}_N\!=\!(\!N\!+\!1\!)!\,v_{N\!+1}(\alpha,\beta,\gamma,
-\alpha\left(N\!+\!1\!+\!(\beta\!+\!\gamma)/2\right),\eta)$
is a constant related to the coefficients in the Taylor series
\ben
\text{HeunC}(\alpha,\beta,\gamma,\delta,\eta,z)=
\sum_{n=0}^\infty v_{n}(\alpha,\beta,\gamma,\delta,\eta)z^n.\hskip .3truecm
\la{HeunC}
\een

These new results are the mathematical basis of the present article.
We derive TSI using the above properties of the confluent Heun function
$\text{HeunC}(\alpha,\beta,\gamma,\delta,\eta,z)$ and show that
the Starobinsky constant is related to its Taylor series coefficients.
This yields a new effective method for the calculation of Starobinsky's
constant for any spin.

Our derivation of the TSI, being simpler, universal and straightforward,
supplements and justifies the usual approach.
It is uniform and valid for both the TAE and TRE.
It extends the TSI to new classes of solutions to the TME, described below.
Thus the known fascinating properties of the standard solutions of the TME
acquire a natural mathematical framework and proper extension.

Moreover, the same derivation yields Teukolsky-Starobinsky like identities
for the Regge-Wheeler equation (RWE) and Zerilli equation (ZE).
To our knowledge, until now the TSI were not known for the RWE and ZE.

Our approach is in accord with the Chandrasekhar
expectation that the Newman-Penrose formalism for perturbations of the Kerr metric
may "...enable us to discover
new classes of identities among the special functions
of mathematical physics when they occur as solutions of Einstein's equations"
and his suggestion to
"...include Teukolsky's functions among "special"
functions of mathematical physics..." \cite{Chandra}.
Today it is clear that for realization of this general idea
we need to use the Heun functions and some of their generalizations \cite{Batic}.

\section{The Exact Solutions to the TAE, TRE and RWE in terms of the Confluent Heun function and the generalized TSI}

The exact local solutions around the singular points $z_\pm=0$ ($\Leftrightarrow z_\mp=1$)
of the three differential equations
listed in the introduction -- the TRE, TAE, and RWE --
can be written in the following common and universal form \cite{Fiziev1}:
\ben
\la{X}
{}_sX^\pm_{\omega,E,m,\{\sigma\}}(z_\pm) =\hskip 4.4truecm\\
=\varrho( z_{{}_\pm})\,e^{{{\alpha z_{{}_\pm}}/ 2}}\, z_{{}_\pm}^{{{\beta}/ 2}}\,
z_{{}_\mp}^{{{\gamma}/ 2}}\,
\text{HeunC}(\alpha,\beta,\gamma,\delta,\eta, z_{{}_\pm}).
\nonumber
\een

The TSI turn out to be a representation of the basic relation (\ref{HHn_maltese})
in terms of the corresponding $\delta_N$-solutions of type (\ref{X}).
Indeed, let us use the short notation $X_N(z_\pm)$, $X^\maltese_N(z_\pm)$
for the $\delta_N$-solutions (\ref{X}) which correspond to the functions
(\ref{HeunCN}), (\ref{AHeunCN}) of argument $z_\pm$ in the place of the factor
$\text{HeunC}(\alpha,\beta,\gamma,\delta,\eta, z_{{}_\pm})$ in Eq. (\ref{X}).
Then, using relations (\ref{AHeunCNparameters})
and $\varrho^{{}_\maltese}=\varrho$,
we obtain from (\ref{HHn_maltese}) the following universal
form of the generalized Teukolsky-Starobinsky identities:
\ben
\la{TSIXN}
\left(z_{{}_{+}}z_{{}_{-}}\right)^{(N+1)/2}\hat D_\pm^{N+1}X_N(z_\pm)=\mathfrak{P}_N X^\maltese_N(z_\pm),
\een
where
\ben
\la{D}
\hat D_\pm\!&=&\!{\frac{d}{dz_\pm}}-{\frac{\alpha}{2}}\mp{\frac{\beta}{2}}\,{\frac{1}{z_\pm}}\pm
{\frac{\gamma}{2}}\,{\frac{1}{z_\mp}}-{\frac{1}{\varrho}}{\frac{d\varrho}{dz_\pm}}.
\een

For compactness, in the right-hand-side of Eqs. \eqref{X}, \eqref{D}
we are using the following abbreviated notation:
$\alpha\!=\!\sigma_\alpha\alpha_\pm$, $\beta\!=\!\sigma_\beta\beta_\pm$, $\gamma\!=\!\sigma_\gamma\gamma_\pm$;
$\{\sigma\}\!=\!\{\sigma_\alpha,\sigma_\beta,\sigma_\gamma\}$, $\sigma_{\alpha,\beta,\gamma}\!=\!\pm 1$.
Thus, we obtain 16 local solutions to any of the TRE, TAE and RWE,
making use of the confluent Heun's function $\text{HeunC}(\alpha,\beta,\gamma,\delta,\eta, z)$
of different argument $z_{+}$ or $z_{-}$ in the right-hand-side of Eq. \eqref{X} and
changing the signs of the parameters $\alpha,\beta,\gamma$.

The difference among the solutions to the above three problems (the TRE, TAE and RWE)
is in the values of the parameters and arguments in their common form \eqref{X}:

1. According to  \cite{Fiziev1}, for the TAE (\ref{angE}) $\varrho(z_{{}_\pm})\!=\!1$ and
\ben
\label{TAEparameters}
\alpha_{\pm}\!&\!=\!&\!\pm 4a\omega,\,\beta_{\pm}\!=\!s\mp m,\,\gamma_{\pm}\!=\!s\pm m,\,\delta_{\pm}=\pm 4s a\omega,\nonumber \\
\eta_{\pm}\!&\!=\!&\!{\frac{m^2\!+\!s^2}{2}}\mp 2s a\omega\!-\!a^2\omega^2\!-\!E;\hskip 2.7truecm \\
z_{+}\!&\!=\!&\!z_{+}(\theta)\!=\!\left(\cos(\theta/2)\right)^2,\, z_{-}\!=\!z_{-}(\theta)\!=\!\left(\sin(\theta/2)\right)^2.\nonumber
\een

2. For the TRE (\ref{radE}) $\varrho(z_{{}_\pm})\!=\!\Delta^{-s/2}$.
Using the quantities  $p\!=\!\sqrt{r_{+}/r_{-}}\!-\!\sqrt{r_{-}/r_{+}}$,
$\Omega_\pm\!=\!{\sqrt{r_{\mp}/r_{\pm}}}\big/\left(r_{+}\!+\!r_{-}\right)$,
$\Omega_a\!=\!\left(\Omega_{+}\!+\!\Omega_{-}\right)/2\!=\!1/(2a)$,
$\Omega_g\!=\!\sqrt{\Omega_{+}\Omega_{-}}\!=\!1/(2M)$
we have:
\begin{subequations}
\label{TREparameters:abcdef}
\ben
\alpha_{{}_{\pm}}\!&\!=\!&\!\pm 2i\omega(r_{+}-r_{-})\!=\!\pm i p\, {{\omega}/{\Omega}_a}, \label{TREparameters:a}\\
\beta_{{}_{\pm}}\!&\!=\!&\! s \pm 2 i \left(m-\omega/\Omega_{\mp}\right)/p,\label{TREparameters:b}\\
\gamma_{{}_{\pm}}\!&\!=\!&s \mp 2 i \left(m-\omega/\Omega_{\pm}\right)/p, \label{TREparameters:c}\\
\delta_{{}_{\pm}}\!&\!=\!&\alpha_{\pm}\left(s-i\omega/\Omega_{g}\right),\label{TREparameters:d}\\
\eta_{{}_{\pm}}\!&\!=\!&\!-E+s^2\!+m^2\!+
{\frac{2m^2\Omega_a^2-\omega^2}{p^2\Omega_a^2}}\,-\nonumber\\
\!&-&\!{\frac{(2m\Omega_a-\omega)^2}{p^2\Omega_g^2}}
-{\frac 1 2}\left({s\!-\!i\,\frac{\omega\,\Omega_{\pm}}{\Omega_a\Omega_g}}\right)^2;\label{TREparameters:e}\\
z_{+}\!&\!=\!&\!{\frac{r\!-\!r_{-}}{r_{+}\!-\!r_{-}}},\,z_{-}\!=\!{\frac{r_{+}\!-\!r}{r_{+}\!-\!r_{-}}}.\label{TREparameters:f}
\een
\end{subequations}

3. For the Schwarzschild metric the discussed results (in terms of the Weyl scalars)
can be obtained from the case of the Kerr metric in the limit $a\to 0$, see \cite{Fiziev1}.
An independent treatment is possible, making use of the RWE.
Then in units $2M=1$ we have \cite{Fiziev1,Fiziev3}:
\begin{subequations}
\label{RWparameters:abcd}
\ben
\varrho \,\, & \!=\! & \, r,\,\, E\,=\,l(l+1),\,\,l=|s|, |s|+1,\dots; \hskip .3truecm\la{RWparameters:a}\\
\alpha_{{}_{\pm}}\!&\!=\!&\pm 2i\omega,\, \beta_{{}_{\pm}}\!=
\!\left\{{\begin{matrix} 2s\\2i\omega\end{matrix}}\right\},
\gamma_{{}_{\pm}}\!=\!\left\{{\begin{matrix} 2i\omega\\2s\end{matrix}}\right\}, \la{RWparameters:b}\\
\delta_{{}_{\pm}}\!&\!=\!&\!\pm2\omega^2,\,\eta_{{}_{\pm}}\!\!=\!\left\{ {\begin{matrix}
-E+s^2\\-E+s^2+2\omega^2 \end{matrix}} \right\}; \la{RWparameters:c}\\
z_{+}&\!\!=\!&\,r,\,\,\,z_{-}=1-r.
\la{RWparameters:d}
\een
\end{subequations}

The knowledge of the 16 local solutions \eqref{X} to the each of the TRE, TAE and RWE
makes it possible to write down their local general solutions as linear combinations
of different pairs of linearly independent particular solutions.
Then we can formulate different boundary problems for the TRE, TAE and RWE \cite{Fiziev1}.
For example, one can impose the standard black hole boundary conditions
on the solutions of the TRE and combine them with the standard regularity requirement
of the solutions to the TAE on the poles $\theta=0,\pi$.
This way we obtain the standard quasinormal modes  of the Kerr black holes
in terms of confluent Heun's functions \cite{Fiziev1}.
Considering polynomial solutions of the TRE and TAE
we are able to describe collimated one-way running waves
of perturbations to the Kerr metric in terms
of confluent Heun's polynomials \cite{Fiziev1}.
One may hope that the proper usage of the singular solutions to the TAE
may help us to describe the creation of relativistic jets by
such collimated running waves, see \cite{Fiziev1,PFDS}.
These important physical problems need a separate careful investigation.

In the present article we are concentrating our efforts on
the mathematical properties of the solutions related with the TSI.
Knowledge of these properties is necessary for further progress in the field.
As we shall see in the next sections,
it is natural to formulate these common general properties
in terms of confluent Heun's functions.

\section{$\delta_N$-confluent Heun's functions and solutions to the TAE, TRE and RWE}

The fulfilment of the $\delta_N$ condition for solutions to the TAE, TRE and RWE
is tightly related to the specific values of the parameters
(\ref{TAEparameters}), (\ref{TREparameters:abcdef}), or (\ref{RWparameters:abcd}).
One can find a detailed general consideration of this problem in \cite{Fiziev1}.

For the TAE the $\delta_N$ condition yields two different cases:
\ben
{}_sN\!+\!1\!=\!2|s|\geq 1,\,|s|\geq 1/2,\,\,\sigma_\alpha\!=\!\sigma_\beta\!=\!\sigma_\gamma\!=\!-\sigma.
\la{sN_Kerr}
\een
\ben
{}_sN_{m}\!+\!1=\!|s|\pm\,m \sigma\geq 1,\,\,\sigma_\alpha\!=\!-\sigma_\beta\!=\!\sigma_\gamma\!=\!-\sigma.
\la{sNm_Kerr_A}
\een

For the TRE the $\delta_N$ condition yields four different cases.
In the first one relations (\ref{sN_Kerr}) are also valid.
Note that there $\sigma=\text{sign}(s)$.
In the other three cases instead of the integer $N$,
the frequencies ${}_s\omega^{\pm}_{N,m,\sigma_\alpha,\sigma_\beta,\sigma_\gamma}$ are fixed:
\ben
{}_s\omega^{+}_{N,m,\mp,\pm,\pm}={}_s\omega^{-}_{N,m,\mp,\pm,\pm}=\pm i\,{\frac {N+1} {4M}},
\la{Im_omega_N_Kerr}
\een
\begin{subequations}\label{omega_N:ab}
\ben
{}_s\omega^{+}_{N,m,\mp,\mp,\pm}={}_s\omega^{-}_{N,m,\mp,\pm,\mp}=\hskip 1.5truecm\label{omega_N:a}\\
=m\Omega_{+}\pm{\frac{i}{4M}}\left(1-{\frac{r_{-}}{r_{+}}}\right)(N+1 \mp s),
\nonumber\\
{}_s\omega^{+}_{N,m,\pm,\mp,\pm}={}_s\omega^{-}_{N,m,\pm,\pm,\mp}=\hskip 1.5truecm\label{omega_N:b}\\
=m\Omega_{-}\pm{\frac{i}{4M}}\left({\frac{r_{+}}{r_{-}}}-1\right)(N+1 \pm s).
\nonumber
\een
\end{subequations}

The $\delta_N$ condition yields  the following cases for the RWE:
\ben
\la{sN_RWE}
{}_sN\!+\!1\!=\!|s|\geq 1,\hskip 2.4truecm\\
\sigma_\alpha\!=\!\sigma_\gamma,\,\sigma_\beta\!=\!-\sigma,\,\,\text{or}
\,\,\,\sigma_\alpha\!=\!\sigma_\beta,\,\sigma
_\gamma\!=\!-\sigma,
\nonumber
\een
\vskip -1truecm
\begin{subequations}\label{RWEomega_N:ab}
\ben
{}_s\omega^{+}_{N\sigma_\beta\sigma_\gamma}\!&\!=\!&\!{\frac {i\sigma_\gamma} 2}(N\!+\!1\!+\!\sigma_\beta s),
\,\sigma_\alpha=-\sigma_\beta;
\hskip .5truecm\label{RWEomega_N:a}\\
{}_s\omega^{-}_{N\sigma_\beta\sigma_\gamma}\!&\!=\!&\!{\frac {i\sigma_\beta} 2}(N\!+\!1\!+\!\sigma_\gamma s),
\,\sigma_\alpha=-\sigma_\gamma.
\hskip .5truecm\label{RWEomega_N:b}
\een
\end{subequations}

Until recently only solutions of a very special kind
were studied and used in the existing literature.
For these special solutions the relations (\ref{sN_Kerr}) take place
and the $\delta_N$-condition is fulfilled automatically
and simultaneously for both the TRE and TAE.

As we have seen, there exist a lot of other interesting solutions to the TRE, TAE and RWE,
listed above, for which the $\delta_N$-condition yields important extra restrictions
on the free parameters of the problem.
We call "$\delta_N$-solution" any solution \eqref{X}
to the TAE, TRE, or RWE, subject to the $\delta_N$ condition \eqref{deltaN}.
In section 4 we will use the term "$\delta_N$-solution" for the solutions \eqref{ZR:b} to the ZE,
having in mind the fulfillment of the $\delta_N$ condition \eqref{deltaN}, too.

Up to now only $\delta_N$-solutions to the TRE and TAE
which obey the relations \eqref{sN_Kerr}  were studied.
The $\delta_N$-solutions, which obey the other relations --
\eqref{sNm_Kerr_A}, for the TAE,
\eqref{Im_omega_N_Kerr}-\eqref{omega_N:ab}, for the TRE
and \eqref{RWEomega_N:ab}, for the RWE,  are new.
Their physical meaning and applications still have to be recovered.
In the next section we derive the generalized TSI, valid for all
$\delta_N$ solutions.

\section{A Novel derivation of the Teukolsky-Starobinsky Identities}
For all eight classes of $\delta_N$-solutions to the TRE, TAE and RWE, described in Sec. III,
one can derive in a uniform way identities of the Teukolsky-Starobinsky type,
starting from the basic form \eqref{TSIXN}.
Relations (\ref{D}), (\ref{TAEparameters}) and (\ref{TREparameters:abcdef})
yield the following explicit representations of the generalized TSI \footnote{To simplify
notation, in Section IV we suppress the indexes
$\omega,E,m,\sigma_\alpha,\sigma_\beta,\sigma_\gamma$ (see \cite{Fiziev1})
of the corresponding quantities.}:

1. In the case of the TAE,
\begin{subequations}\la{TSI_Nm_Kerr_A:a,b,c,d}
\ben
\big(\sin\theta\big)^{N+1}\left({\frac{\hat d_{\theta,\pm}}{\sin\theta}}\right)^{\!\!N+1}\!\!S_{N}^\pm(\theta)
\!=\!\mathfrak{C}_N S_{N}^{\pm\maltese}(\theta),
\hskip 1.9truecm\la{TSI_N_Kerr_A:a}\\
\hat d_{\theta,\pm}\!=\!{\frac{d}{d\theta}}\!+\!\sigma_\alpha a\omega\sin\theta+
{\frac{(\sigma_\beta\!-\!\sigma_\gamma)s\!\mp\!(\sigma_\beta\!+\!\sigma_\gamma)m}{2\sin\theta}}\!\mp\hskip .9truecm\nonumber\\
\mp{\frac{(\sigma_\beta\!+\!\sigma_\gamma)s\!\mp\!(\sigma_\beta-\sigma_\gamma)m}{2}}\cot\theta,
\hskip 3.truecm\la{TSI_N_Kerr_A:b}\\
S_{\,N}^\pm(\theta)=X_N(z_\pm),
\hskip 5.7truecm\la{TSI_N_Kerr_A:c}\\
\mathfrak{C}_N=(\pm 1)^{N+1}\,\mathfrak{P}_N.
\hskip 5.45truecm\la{TSI_N_Kerr_A:d}
\een
\end{subequations}

Disentangling the differential operators in Eq. (\ref{TSI_N_Kerr_A:a}),
we can represent it in a form similar to Eq. \eqref{TSI_A}:
\ben
\prod\limits_{\overleftarrow{k=0}}^{N+1}\left(\hat d_{\theta,\pm}-k\cot\theta\right)\!S_{N}^\pm(\theta)
\!=\!\mathfrak{C}_N S_{N}^{\pm\maltese}(\theta).
\la{TSI_Nm_Kerr_A}
\een
Here an ordered operator product is used.
The arrow indicates the operator ordering and points in the direction of
the increase of the integer $k$.

2. In the case of the TRE,
\begin{subequations}\la{TSI_Nm_Kerr_R:a,b,c,d}
\ben
\Delta^{(N+1)/2}\left(\hat d_{r,\pm}\right)^{N+1}R_{\,N}^\pm(r)=\mathfrak{D}_N R^{\pm\maltese}_{\,N}(r),
\hskip 1.6truecm\la{TSI_Nm_Kerr_R:a}\\
\hat d_{r,\pm}\!=\!{\frac d {dr}}\!-i\sigma_\alpha\omega\pm \hskip 5.05truecm \nonumber\\
\pm\, i\, {\frac{\sigma_\gamma(m\!-\!\omega/\Omega_\pm)}{p\,(r\!-\!r_{+})}}
\mp i\, {\frac{\sigma_\beta(m\!-\!\omega/\Omega_\mp)}{p\,(r\!-\!r_{-})}},
\hskip 1.15truecm\la{TSI_Nm_Kerr_R:b}\\
R_{\,N}^\pm(r)\!=\!(r\!-\!r_{+})^{s(1\!-\!\sigma_\gamma)/2}(r\!-\!r_{-})^{s(1\!-\!\sigma_\beta)/2}X_N(z_\pm),
\hskip .65truecm\la{TSI_Nm_Kerr_R:c}\\
\mathfrak{D}_N=(\pm 1)^{N+1}(-1)^{(N+1)/2}\,\mathfrak{P}_N.
\hskip 3.25truecm\la{TSI_Nm_Kerr_R:d}
\een
\end{subequations}

It is easy to check that in the case  of Eqs. (\ref{sN_Kerr}) relations
(\ref{TSI_Nm_Kerr_A:a,b,c,d}) and (\ref{TSI_Nm_Kerr_R:a,b,c,d}) produce
the standard TSI \eqref{TSI_A} and \eqref{TSI_R}
with  Starobinsky's constants $\mathfrak{C}_N$ and $\mathfrak{D}_N$.

To reach this result one has to apply the new nontrivial hidden symmetry
of the parameters of $\delta_N$-confluent Heun's functions,
considered as functions of the spin-weight $s$
in the special case $N\!+\!1\!=\!2|s|$. Then:
\ben
\{\alpha_{\pm}^{{}_\maltese}(s),\beta_{\pm}^{{}_\maltese}(s),\gamma_{\pm}^{{}_\maltese}(s),
\delta_{\pm}^{{}_\maltese}(s),\eta_{\pm}^{{}_\maltese}(s)\}\!=\!
\hskip 1.8truecm \nonumber\\
=\!\{\alpha_{\pm}(-s),\beta_{\pm}(-s),\gamma_{\pm}(-s),\delta_{\pm}(-s),\eta_{\pm}(-s)\}.
\hskip .3truecm \la{maltese_symmetry}
\een
Equation (\ref{maltese_symmetry}) follows from (\ref{AHeunCNparameters}),
(\ref{TAEparameters}) and (\ref{TREparameters:abcdef}), if
$N\!+\!1\!=\!2|s|$. As a result, in the case (\ref{sN_Kerr})
$X^\maltese_N(s,z_\pm)=X_N(-s,z_\pm)$ is a solution
to {\em the same} TRE or TAE but with spin-weight $(-s)$.
The last fact was discovered for the first time in \cite{Teukolsky},
as a special property of the particular solutions to the TRE or TAE,
considered there.
As seen, we obtain a natural explanation of this phenomenon
using the symmetry properties of $\delta_N$-confluent Heun's functions.

Now it remains only to point out that

A) From Eqs. \eqref{TSI_Nm_Kerr_A:a,b,c,d} for
$\sigma_\alpha\!=\!\sigma_\beta\!=\!\sigma_\gamma\!=\!-\sigma$ and
$n=s-k$ one obtains
$$\hat d_{\theta,+}-k\cot\theta={}_{n}\mathcal{L},\,\,
S_{N}^{+}(\theta)={}_{\!+s} S(\theta),\,\,
S_{N}^{+\maltese}(\theta)={}_{\!-s} S(\theta)$$
when $\sigma=+1$; or
$$\hat d_{\theta,+}-k\cot\theta={}_{n}\mathcal{L}^\dag,\,\,
S_{N}^{+}(\theta)={}_{\!-s} S(\theta),\,\,
S_{N}^{+\maltese}(\theta)={}_{\!+s} S(\theta)$$
when $\sigma=-1$. Hence, for these particular values of the parameters
Eq. \eqref{TSI_Nm_Kerr_A} coincides with the standard angular TSI \eqref{TSI_A}.

B) From Eqs. \eqref{TSI_Nm_Kerr_R:a,b,c,d} for
$\sigma_\alpha\!=\!\sigma_\beta\!=\!\sigma_\gamma\!=\!-\sigma$
one obtains
$$\hat d_{r,+}={}_0\mathcal{D},\,\,
R^{+}_{\,N}(r)=\Delta^s\,{}_{+s}R(r),\,\,
R^{+\maltese}_{\,N}(r)={}_{-s}R(r),$$
and $\mathfrak{D}_N={}_s\mathfrak{D}$ when $\sigma=+1$; or
$$\hat d_{r,+}={}_0\mathcal{D}^{\dag},\,\,
R^{+}_{\,N}(r)={}_{-s}R(r),\,\,
R^{+\maltese}_{\,N}(r)=\Delta^s\,{}_{+s}R(r),$$
and $\mathfrak{D}_N={}_s\mathfrak{D}^*$ when $\sigma=-1$. Hence, for these particular values of the parameters
Eq. \eqref{TSI_Nm_Kerr_R:a} coincides with the standard radial TSI \eqref{TSI_R}.

To our knowledge, the TSI for the other seven cases --
Eqs. \eqref{sNm_Kerr_A}, \eqref{Im_omega_N_Kerr}, \eqref{omega_N:a}, \eqref{omega_N:b},
\eqref{sN_RWE}, \eqref{RWEomega_N:a}, and \eqref{RWEomega_N:b}
have not been studied up to now. In particular, relation (\ref{TSI_Nm_Kerr_R:c})
extends to all cases (\ref{Im_omega_N_Kerr}), (\ref{omega_N:ab})
the transformation invented for the case (\ref{sN_Kerr}) in
\cite{Teukolsky,ChandraMT}.

3. For the RWE (in units $2M=1\Rightarrow\Delta_0=r(r-1)$):
\begin{subequations}\la{TSI_RWE:a,b,c,d}
\ben
\Delta_0^{(N+1)/2}\left(\hat d_{r,0,\pm}\right)^{N+1}R_{N,0}^\pm(r)=\mathfrak{D}_{N,0} R^{\pm\maltese}_{N,0}(r),
\hskip 1.truecm\la{TSI_RWE:a}\\
\hat d_{r,0,\pm}\!=\!{\frac d {dr}}\!-i\sigma_\alpha\omega-\hskip 4.65truecm\nonumber\\
-{\frac{\sigma_\beta\!-\!\sigma_\gamma}{2}}(s\!+\!i\omega)\mp{\frac{\sigma_\beta\!+\!\sigma_\gamma}{2}}(s\!-\!i\omega),
\hskip 1.2truecm\la{TTSI_RWE:b}\\
R_{N,0}^\pm(r)=r^{-1}\,X_N(z_\pm),
\hskip 4.25truecm\la{TSI_RWE:c}\\
\mathfrak{D}_{N,0}=(\pm 1)^{N+1}\,\mathfrak{P}_N.
\hskip 4.6truecm\la{TSI_RWE:d}
\een
\end{subequations}

For the TSI (\ref{TSI_RWE:a,b,c,d}) we have $N\!+\!1\!=\!|s|$.

4. The case of the  ZE is slightly different.
In it we may proceed in the following way.

The solutions $Z_{0}^\pm(r)$ to the Zerilli equation are known to be simply related
with the solutions $R_{0}^\pm(r)$ to the RWE \cite{ChandraMT}.
We write down the corresponding relations (in units $2M=1$)
in the following form (see the appendix):
\begin{subequations}\la{ZR:a,b}
\ben
i\bar\omega R_{0}^\pm(r)&=&\left(1-{\frac 1 r}\right)e^{\psi(r)}{\frac d {dr}}\left(e^{-\psi(r)}Z_{0}^\pm(r) \right),
\hskip 0.9truecm\la{ZR:a}\\
i\bar\omega Z_{0}^\pm(r)&=&\left(1-{\frac 1 r}\right)e^{-\psi(r)}{\frac d {dr}}\left(e^{\psi(r)}R_{0}^\pm(r) \right).
\hskip 0.5truecm\la{ZR:b}
\een
\end{subequations}

Let us introduce the notation $ Z^{\pm}_{N,0}(r)$ for $\delta_N$ solutions to the ZE
and $ Z^{\pm\maltese}_{N,0}(r)$ for corresponding associate solutions, and the new operator
${}_{{}_Z}\hat d_{r,0,\pm}=\hat d_{r,0,\pm}+\Delta_0^{-1}$.
Then after substitution of \eqref{ZR:a} into
Eq. \eqref{TSI_RWE:a} we obtain the following nonstandard Teukolsky-Starobinsky-like
identities for the solutions to the ZE:
\ben\la{TSI_Z}
\Delta_0^{\frac{N+1}2}\left({}_{{}_Z}\hat d_{r,0,\pm}\right)^{\!N+1}
\left(e^{\psi(r)}{\frac d {dr}}\left(e^{-\psi(r)}Z_{N,0}^\pm(r) \right)\right)=\nonumber\\
=\mathfrak{D}_{N,0} \left(e^{\psi(r)}{\frac d {dr}}\left(e^{-\psi(r)}Z_{N,0}^{\pm\maltese}(r) \right)\right).
\hskip 1.5truecm
\een

\section{A new method for calculation of Starobinsky's constants $\mathfrak{C}_N$ and $\mathfrak{D}_N$}

Equations (\ref{TSI_Nm_Kerr_R:d}), (\ref{TSI_N_Kerr_A:d}) and (\ref{TSI_RWE:d}) show
that in all cases the corresponding Starobinsky constants
coincide (up to known numerical factors)
with the coefficients $v_{n}(\alpha,\beta,\gamma,\delta,\eta)$
in the Taylor series expansion (\ref{HeunC}) of $\delta_N$-confluent Heun's function
evaluated for proper values of the parameters.
These coefficients can be determined by the three-terms
recurrence relation \cite{Fiziev2,Heun}:
\ben
A_{n}v_{n}=B_{n}v_{n-1}+C_{n}v_{n-2},\,\,\,\,n=1,\dots,\infty,
\la{recurrence}
\een
with the initial condition $v_{-1}=0,\,\,v_{0}=1$.
For $\delta_N$-confluent Heun's functions in (\ref{recurrence}) we have coefficients
\ben
\la{rec_coeff}
A_{n}&=&1+{\frac{\beta}{n}},\\
B_{n}&=&1-{\frac{\varkappa+1}{n}}+
{\frac{\varkappa-\mu}{n^2}},\nonumber\\
C_{n}&=&{\frac{\alpha}{n^2}}\left(n-N-2\right).\nonumber
\een

Relations (\ref{recurrence}), (\ref{rec_coeff}) give a very efficient
new method for calculation of Starobinsky's constant
for all values of the spin $|s|$, confirming the known results and
easily solving the problem for all cases (\ref{sN_Kerr})-(\ref{RWEomega_N:ab})
of validity of the $\delta_N$-condition.

\section{Conclusion}
In the present article, we showed that TSI exist for a large class of physical problems
when the background vacuum metric is of Petrov type D.
We constructed explicitly the TSI for different perturbations of any spin to
the Kerr metric, as well as to the Schwarzschild one.
Our derivation of TSI is valid for complex values of the separation constants $\omega$ and $E$,
as well as for complex values of all other parameters.
The obtained results can be applied to both the black holes and naked singularities.

The exact solutions to the TRE, TAE, RWE and ZE were briefly described
in terms of confluent Heun's functions.
One can find more details in the articles \cite{Fiziev1}.
Having in hands these explicit solutions we are able

1. To write down the corresponding general solutions.

2. To formulate different boundary problems.

3. To derive all properties of the solutions
using explicitly the properties of Heun's functions.

In particular this article accomplished the following aims:

i) We found the TSI and all their possible generalizations have a common origin --
the recently unveiled property \eqref{HHn_maltese} of the confluent Heun function,
related with the $\delta_N$ condition (\ref{deltaN}), see \cite{Fiziev2}.
As a result of this condition, the multiple derivative of order $(N+1)$
of a given $\delta_N$-confluent Heun's function becomes
proportional to another (associated) confluent Heun's function.
The coefficient is precisely $(N+1)!$ times the $(N+1)$-th coefficient
in the Taylor series expansion of the first function.
This algebraic property is universal and not related to the usually considered specific
boundary problems, nor with other properties of their solutions like regularity, integrability,
etc.
Using the relation of the exact solutions at hand with the confluent Heun's function,
one can express this property directly in terms of the very solutions.
Thus one obtains the TSI for the TRE, TAE, RWE and ZE in a uniform and universal way.

ii) We discussed all classes of $\delta_N$ solutions
to the above problems.
For the specific class of $\delta_N$-solutions, defined by relation (\ref{sN_Kerr}),
the novel symmetry (\ref{maltese_symmetry}) of the parameters of
$\delta_N$-confluent Heun's functions (\ref{HeunCN}) ensures that its multiple
derivative of order $(N+1)$ is a solution to the {\em same} Heun equation.
Thus, this symmetry yields a second solution to the same TME equation.
As a result, we obtain the known pair of solutions to the TME with the opposite signs
of their spin-weights $(\pm s)$ \cite{Teukolsky}.

iii) We extended the familiar results for TSI to all
possible cases on a firm and general mathematical basis.
Thus, using the $\delta_N$ condition we found the possible
generalizations of the TSI.

iv) We made transparent the origin of some of the properties
of the solutions to the TRE, TAE, RWE and ZE,
relating them with the specific form of the parameters in confluent Heun's functions,
which solves these equations.
For example, such unexpected property is the symmetry \eqref{maltese_symmetry},
which is specific for large class of solutions to TRE and TAE. This
symmetry does not take place for all confluent $\delta_N$-Heun's functions.

v) Using the recurrence relation for the Taylor series expansion of the confluent Heun's function
we proposed a new efficient and universal method for the calculation of Starobinsky's constants.

Thus, the confluent Heun function was proven to be
an adequate tool for solving of the above problems.
The usage of this function gives a natural treatment
and deeper understanding of the Teukolsky-Starobinsky identities,
as well as of the solutions to the Teukolsky equations.
The results, presented here, are a basis for future physical applications.

\begin{acknowledgments}
I am deeply grateful to A.~Starobinsky and to G.~Alekseev for the useful discussions,
as well as to the unknown referees for the very helpful suggestions.

I am thankful to Bogolubov Laboratory of Theoretical Physics, JUNR, Dubna, Russia for the hospitality
and good working condition during my stay there in the summer of 2009.

This article was supported by the National Scientific Found of the Bulgarian Ministry of
Education and Science under Contracts No. DO-1-872, No. DO-1-895 and No. DO-02-136,
and by the Sofia University Foundation "Theoretical
and Computational Physics and Astrophysics."
\end{acknowledgments}

\appendix
\section{On the relation between the solutions to the RWE and ZE}
The relation between the solutions to the RWE and ZE were discovered by Chandrasekhar,
see for example \cite{ChandraMT} and the references therein.
We slightly modify the form of these relations using
the notation $Z_{0}^\pm(r)$ for solutions to the Zerilli equation and $R_{0}^\pm(r)$
for solutions to the RWE. These functions have additional constant phases
in comparison with the functions $Z^{(\pm)}(r)$, used in \cite{ChandraMT}
(Note that the meaning of the signs '$\pm$' in our notation is completely different
see \cite{Fiziev1}.) Our choice
of these additional phases ensures the appearance of the {\em common} factor $i\bar\omega$
simultaneously in the two of the Eqs. \eqref{ZR:a,b}.

As usual, here we consider only the ZE for gravitational waves with spin $|s|=2 \Rightarrow N+1=2$.
Then in units $2M=1$ we obtain
\ben
e^{\psi(r)}&=&|r-1|^{\omega_l}e^{\omega_l r}\left(1+{\frac 3 {(l-1)(l+2)r}}\right)^{-1}\!\!,\hskip .9truecm\\
\omega_l&=&{\frac 1 6}(l-1)l(l+1)(l+2),\,\,\,l=2,3,\dots\, ,\nonumber\\
\bar\omega&=&\sqrt{\omega^2+\omega_l^2}\neq 0\,\,\,\,\,\text{for}\,\,\,\,\, \omega\neq \pm i \omega_l.\nonumber
\la{psi}
\een
The Eqs. \eqref{ZR:a,b} do not relate the solutions $Z_{0}^\pm(r)$  and  $R_{0}^\pm(r)$
when $\omega = \pm i \omega_l$. Instead, in this case one obtains from
the Eqs. \eqref{ZR:a,b} the well known algebraically special solutions
in the following explicit quasipolynomial form:
\begin{subequations}\la{RZsol:a,b}
\ben
R_{0}^\pm(r)&=&\text{const}\times e^{+\psi(r)},
\la{ZRsol:a}\\
Z_{0}^\pm(r)&=&\text{const}\times e^{-\psi(r)}.
\la{ZRsol:b}
\een
\end{subequations}
Thus we see that the case of the algebraically special solutions \eqref{RZsol:a,b} is a degenerate one.

The system of the ordinary differential equations \eqref{ZR:a,b}
can be easily split to the decoupled RWE and ZE, written in the form
\begin{subequations}\la{ZREq:a,b}
\ben
R^{\prime\prime}+\bar\omega^2 R&=&\left(-\psi^{\prime\prime}+{\psi^\prime}^2\right)R,
\hskip 0.9truecm\la{ZREq:a}\\
Z^{\prime\prime}+\bar\omega^2 Z&=&\left(+\psi^{\prime\prime}+{\psi^\prime}^2\right)Z.
\hskip 0.5truecm\la{ZREq:b}
\een
\end{subequations}
Here the prime denotes differentiation with respect to the tortoise coordinate $r_*=r+\ln|r-1|$.
This splitting reflects the physical independence of the axial and polar perturbations
to the Schwarzschild metric.


\begin{thebibliography}{}
%
\bibitem{Teukolsky} S.~A.~Teukolsky, PRL {\bf 29}, 1114 (1972).
                    S.~A.~Teukolsky, ApJ {\bf 185}, 635 (1973).
                    W.~H.~Press, S.~A.~Teukolsky, ApJ {\bf 185}, 649 (1973).
                    S.~A.~Teukolsky, W.~H.~Press, ApJ {\bf 193}, 443 (1974 ).
%
\bibitem{Starobinsky} A. A. Starobinskiy, Zh. Exp. Theor. Fiz. {\bf 64}, 48 (1973) (transl. Sov. Phys. JETP {\bf 37} 28 (1973));
                      A. A. Starobinskiy, S. M. Churilov ,  Zh. Exp. Theor. Fiz. {\bf 65}, 3 (1973) (transl. Sov. Phys. JETP {\bf 38} 1 (1973)).
%
\bibitem{Chandra}   S.~Chandrasekhar,  Proc. R. Soc. London {\bf A348 }, 39 (1976);
                    Proc. R. Soc. London {\bf A349 }, 1  (1976);
                    Proc. R. Soc. London {\bf A358 }, 405 (1978);
                    Proc. R. Soc. London {\bf A358 }, 421 (1978);
                    Proc. R. Soc. London {\bf A358 }, 441 (1978);
                    Proc. R. Soc. London {\bf A365 }, 425 (1979);
                    Proc. R. Soc. London {\bf A365 }, 453 (1979);
                    Proc. R. Soc. London {\bf A372 }, 475 (1980).
%
\bibitem{ChandraMT}   S.~Chandrasekhar,  {\em The Mathematical Theory of Black
                    Holes}, Oxford University Press, Oxford, 1983.
%
\bibitem{other}     P. L. Chrzanowski, Phys. Rev. D{\bf 11}, 2042 (1975).
                    R. M. Wald, Phys. Rev. Lett. {\bf 41}, 203 (1978).
                    R. M. Wald, Gen. Rel. Grav, {\bf 11}, 321 (1979).
                    G. Silva-Ortigoza, J. Math. Phys. {\bf 42}, 368 (2001)
%
\bibitem{TSI}       U. Khanal, Phys. Rev. D{\bf 32}, 879 (1985).
                    G. F. Torres del Castillo, J. Math. Phys. {\bf 29}, 2078 (1988).
                    G. F. Torres del Castillo, J. Math. Phys. {\bf 30}, 446 (1989).
                    E. G. Kalnins, W. Miller, Jr., G. C. Williams, J. Math. Phys. {\bf 30}, 2925 (1989).
                    G. Silva-Ortigoza, Revista Mexicana de F\'isica {\bf 40}, 730-737  (1994).
                    G. Silva-Ortigoza,  Class. Quant. Grav. {\bf 14},  795-804 (1997).
                    S.~Mano, E.~Takasugi, Progr.Theor. Phys. {\bf 97}, 213 (1997).
%
\bibitem{TSIapplications}  C. M. Chambers, I. G. Moss, Class. Quant. Grav. {\bf 11}, 1035-1054 (1994).
                           L. Barack, A. Ori, Phys. Rev. D{\bf 60}, 124005 (1999).
                           Shuang-Qing Wu, Mu-Lin Yan, Phys. Rev. D{\bf 69}, 044019 (2004).
%
\bibitem{VCMC}    V. Cardoso, M. Cavaglia, Phys. Rev. D{\bf 74}, 024027 (2006).
                  G. Dotti, R. Gleiser, J. Pullin, Phys. Lett. B{\bf 644}, 289-293 (2007)
                  G. Dotti, R. J. Gleiser, J. Pullin, I. F. Ranea-Sandoval, H. Vucetich,
                  Int. J. Mod. Phys. A{\em 24}, 1578-1582 (2009)
%
\bibitem{STARA}     S. K. Chakrabarti, P. S. Joshi, Int.J.Mod.Phys. D{\ bf 3}  647-651 (1994).
                    T. P. Singh, Gen. Rel. Grav. {\bf 30}, 1563-1567, 1998.
                    T. Harada, H. Iguchi, Ken-ichi Nakao, Phys. Rev. D{\em 61}, 101502  (2000).
                    H. Iguchi, T. Harada, Class. Quant. Grav. {\em 18}  3681-3700 (2001).
%
\bibitem{QNM}       V.~Ferrari,  in {\em Proc. of 7-th Marcel Grossmann Meeting},
                    ed, R. Ruffini, M. Kaiser, Singapore, World Scientific, 1995;
                    and  in {\em Black Holes and Relativistic
                    Stars}, ed. R.~Wald, Univ. Chicago Press, Chicago 1998.
                    K.~D.~Kokkotas, B.~G.~Schmidt,  Living Rev. Relativity {\bf 2}, 2 (1999).
                    H-P.~Nollert,  Class. Quant. Grav. {\bf 16}, R159,
                    (1999),  V.~Ferrari, L.~Gualtieri, Class. Quant. Grav. {\bf 40} 945-970 (2008).
                    E. Berti, V. Cardoso, A. O. Starinets, {Quasinormal modes of black holes and black branes},
                    Classical Quantum Gravity 26, 163001 (2009).
%
\bibitem{TME_Heun}  G.~Marcilhacy, Lett. Nuovo Cim. {\bf 37}, 300-302 (1983).
                    J.~Blandin, R.~Pons, G.~Marcilhacy, Lett. Nuovo Cim. {\bf 38}, 561-567 (1983).
                    D.~V.~Galtsov, A.~A.~Ershov, Izv. Vuzov Fiz. {\bf 32}, 13-18 (1989) (transl. Soviet Phys. Jour. {\bf 32}, 764 (1990)).
                    H.~Suziki, E.~Takasugi, H.~Umetsu, Progr. Theor. Phys. {\bf 100}, 491 (1998).
                    D.~Batic, H.~Schmid, J. Math. Phys. {\bf 48}, 042502 (2007).
%
\bibitem{Fiziev1} Fiziev~P.~P., {\em Classes of Exact Solutions to Regge-Wheeler and Teukolsky  Equations}, gr-qc/0902.1277
                                {\em Classes of Exact Solutions to the Teukolsky  Master Equation}, gr-qc/0908.4234
%
\bibitem{Fiziev2} Fiziev~P.~P., {\em Novel relations and new properties of confluent Heun's functions and their derivatives of arbitrary order},
                  arXiv:0904.0245 [math-ph]
%
\bibitem{Fiziev3} Fiziev~P.~P., Class. Quant. Grav., {\bf 23}, 2447-2468 (2006).
                  Fiziev~P.~P., Jour. Phys. Conf. Ser. {\bf 66}, 012016 (2007).
%
\bibitem{Heun}      K.~Heun,  Math. Ann. {\bf 33}, 161 (1889).
                    H.~Bateman, A.~Erd\'elyi,  {Higher Transcendental
                    Functions}, Vol. 3, McGraw-Hill Comp., INC, New York 1955.
                    A.~Decarreau, M.~Cl.~Dumont-Lepage, P.~Maroni,
                    A.~Robert, A.~Roneaux,  Ann. Soc. Bruxelles, {\bf
                    92}, 53, (1978).
                    A.~Decarreau, P.~Maroni,A.~Robert,  Ann. Soc. Bruxelles, {\bf
                    92}, 151 (1978).
                    {Heun's Differential Equations}, ed. A.~Roneaux,
                    Oxford Univ. Press, New York 1995.
                    S.Y. Slavyanov and W.Lay,  "Special Functions,
                    A Unified Theory Based on Singularities",
                    Oxford Mathematical Monographs, New York 2000.
                    R.~S.~Maier, {\em The 192 Solutions of Heun
                    Equation}, math.CA/0408317.
%
\bibitem{PFDS}      P.~P.~Fiziev and D.~R.~Staicova, Bulg. Astrophysical J. {\bf 11}, 3  (2009), arXiv:0902.2408;
                                                     Bulg. Astrophysical J. {\bf 11}, 13 (2009), arXiv:0902.2411.
%
\bibitem{Batic}   D.~Batic, H.~Schmid, M.~Winklmeier, J. Phys. A: MAth. Gen. {\bf 39}, 12559-12564 (2006).
                  D.~Batic, M.~Sandoval, Cent.Eur. J. Phys.,
                  {\em The hypergeneralized Heun equation in QFT in curved space-times}, arXiv:0805.4399v1 [gr-qc]
%
\end{thebibliography}
\end{document}